\documentclass[twocolumn, fleqn]{svjour2}    
\smartqed  
\usepackage{graphicx}
\usepackage{mathptmx}      
\usepackage{amsmath}
\usepackage{natbib}
\newcommand{\bm}[1]{\mbox{\boldmath$#1$\unboldmath}}

\begin{document}

\title{Techniques for Generating Centimetric Drops in Microgravity \\
and Application to Cavitation Studies
}
%
\author{P. Kobel  \and D. Obreschkow \and A. de Bosset \and N. Dorsaz \and M. Farhat}
%
%
\institute{P. Kobel  \and D. Obreschkow \and A. de Bosset \and N. Dorsaz \and M. Farhat
           \at   Laboratoire des Machines Hydrauliques, EPFL, 1007 Lausanne, Switzerland \\
            Tel.: +41-21-6932505 \\
            Fax: +41-21-6932551 \\
          \and
           P. Kobel  \at
              Max-Planck Institut f\"{u}r Sonnensystemforschung, Max-Planck-Stra\ss e 2, 37191 Katlenburg-Lindau, Germany \\
            \email{philippe.kobel@a3.epfl.ch}
           \and
           D. Obreschkow \at
              Astrophysics, Department of Physics, University of Oxford, Keble Road, Oxford, OX1 3RH, UK
            \and
           N. Dorsaz \at
              Institut Romand de Recherche Num\'{e}rique en Physique des Mat\'{e}riaux, EPFL, 1015 Lausanne, Switzerland
}
\date{Received: date / Accepted: date}
%
%
\maketitle

\begin{abstract}
This paper describes the techniques and physical parameters used to produce stable centimetric water drops in
microgravity, and to study single cavitation bubbles inside such drops (Parabolic Flight Campaigns, European
Space Agency ESA). While the main scientific results have been presented in a previous paper, we shall herein provide
the necessary technical background, with potential applications to other experiments. First, we present an
original method to produce and capture large stable drops in microgravity. This technique succeeded in generating
quasi-spherical water drops with volumes up to 8 ml, despite the residual g-jitter. We find that the equilibrium
of the drops is essentially dictated by the ratio between the drop volume and the contact surface used to capture
the drop, and formulate a simple stability criterion. In a second part, we present a setup for creating and
studying single cavitation bubbles inside those drops. In addition, we analyze the influence of the bubble size
and position on the drop behaviour after collapse, i.e. jets and surface perturbations.

\end{abstract}

\section{Scientific Background}\label{background}

A large spherical liquid body in absence of gravitational forces is a paradigmatic system to study a variety of
physical phenomena. Yet the control of its position and stability in a residual acceleration environment is
a delicate task. Acoustic positioning techniques, developped for the Drop Physics Module (DPM, flown onboard
space shuttle Columbia mission STS-73), overcame this problem while allowing the manipulation of perfectly
isolated centimeter-sized levitating drops \citep[described in detail by][]{Croon 93}. This complex technique has
proved fruitful for dynamical studies such as the equilibrium shapes of rotating drops \citep{Wang 86}, or the
rheological properties of oscillating drops with adsorbed surfactants \citep{Apfel 97, Holt 97, Chen 98}.
However, many applications do not require the direct manipulation of free-floating drops, and can be successfully
achieved with a captured drop. Particular instances are the study of free surface phenomena, e.g. Faraday waves
\citep{Faraday 31, Holt 96}, thermocapillary flows \citep{Treuner 95}, or liquid jets \citep{Rayleigh 79}. For
these purposes, a much simpler system could be used, such as an injector tube able to attach and stabilize the
drop. We designed and implemented such a drop generation system, and used the ensuing centimetric drops as nearly
isolated spherical volumes in which single cavitation bubbles may be investigated.

The behaviour of cavitation bubbles --depressurized vapor bubbles appearing naturally in many industrial
systems-- has been demonstrated to be strongly dependent on the liquid geometry and the nature of nearby
surfaces, as both impose boundary conditions on the pressure field of the fluid \citep[as reviewed in][]{Brennen
95}. In particular, these geometrical conditions influence the violent emission of liquid jets and shock waves by
collapsing bubbles, identified as the main sources of cavitation erosion in hydraulic machinary \citep{Isselin
98, Shima 97}. Therefore recent fundamental studies focussed on the dynamics of bubbles and associated phenomena
next to flat \citep{Brujan 02} and curved \citep{Tomita 02} rigid surfaces, or flat free surfaces \citep{Robinson
01, Pearson 04}. However, on-ground investigations inside static water flasks are restricted to probe \emph{flat}
free surfaces due to gravity, and the water volume can hardly be isolated. To extend these possibilities, we used
a microgravity environment to produce spherical water drops, inside which we studied cavitation bubbles. It
should be emphasized that our experiment did not aim at investigating the direct effects of microgravity on
cavitation, but used microgravity as a way to produce a spherical water drop of large size, which is crucial to
vary physical parameters such as the size and position of the bubble within the drop (see Sect. \ref{jets and
drop_stability}). Moreover, this drop geometry presents unique properties influencing the bubble dynamics: a
\emph{spherical}, \emph{closed} free surface and an \emph{isolated} volume.

Our experiment was flown on parabolic flights, offering sequences of microgravity phases lasting for about 20
seconds (ESA 42nd microgravity research campaign 2006, 8th student parabolic flight campaign 2005). A variety of
scientific outcomes has been presented by \citet{F&S 06}, including (1) double-jet formation, (2) shock
wave-induced secondary cavitation and (3) shorter collapse time than in extended water volumes.

This paper aims at presenting in detail the intrumentation designed for the creation of the drop and the
generation of the bubble, as well as the physical parameters influencing the performance of the experiment. It is
structured in two main parts: the first part (Sect. \ref{drop part}) addresses the generation of a centimetric
drop and its stability in the residual g-jitter, while the second part (Sect. \ref{cavitation part}) describes
the setup used to create and observe cavitation bubbles inside the drop. The latter section also includes an
analysis of the drop state (liquid jets emission, drop stability) after bubble collapse with respect to the
bubble position and size. To conclude, a brief summary is given in Sect. \ref{summary}, including some
prospects of the drop generation system.

\section{Generation of Centimetric Drop in Microgravity}\label{drop part}
\subsection{Drop Generation System}

\begin{figure}[h]
  \includegraphics[width=\columnwidth]{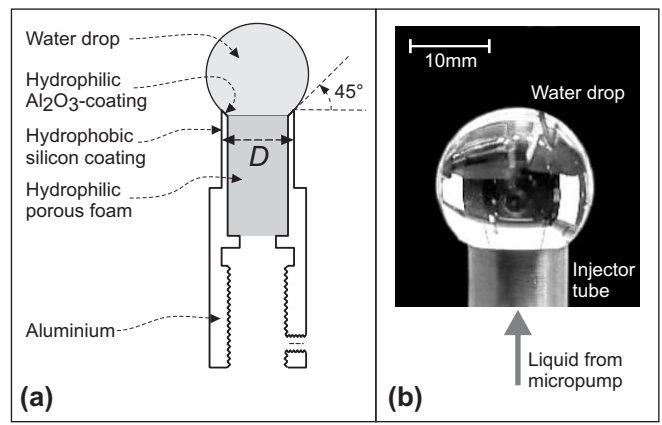}
  \caption{(a) Vertical cross section of the injector tube.
  (b) Quasi-spherical drop in microgravity}
  \label{fig_tubedrop}
\end{figure}

During each microgravity phase (20 s duration), \emph{one} captured centimetric drop was created as follows.
Distilled water was smoothly expelled through a custom-designed Aluminium \emph{injector tube} (Fig.
\ref{fig_tubedrop} (a)) by a step-motor micropump (APT Instruments SP200 peristaltic pump). The injector tube was
filled with hydrophilic porous foam for two purposes. First, to guarantee a homogeneous laminar flow and thereby
control the actual drop volume during its formation. Second, to favor the attachment of the drop by water
capillary cohesive forces through the tube foam interface. In addition, the top edge of the tube was trimmed at
an angle of $45^\circ$ in order to increase the effective contact surface with the drop. This inclined edge was
coated with hydrophilic aluminium oxyde Al$_2$O$_3$ to ensure efficient contact, while the tube exterior was
coated with hydrophobic silicon to prevent the water from flowing down. Four injector tubes with outer diameter
$D=$ 16, 12, 9 and 7 mm have been used (see Fig. \ref{fig_tubedrop} (a)). They were interchanged according to the
water drop volume to be produced, as larger contact surfaces allow to stabilize larger volumes (cf Sect.
\ref{drop_equilibrium}).

The micropump was programmed by an external computer, which triggered the pumping start, controlled its flux and
its duration. The micropump was automatically started whenever the gravity level dropped below 0.05 g for
more than 1 s (see Sect. \ref{general_setup}). The flux was set to 0.6 ml/s and reduced to 0.3 ml/s during the
two--to--three last seconds of the pumping process. This flux reduction aimed at slowing down the drop motion and
avoiding subsequent drop oscillations. Varying the pumping time allowed to adjust the drop size. To precisely
control the drop volume, the tube was filled with water up to its top foam interface prior to start the pumping.
Thus, drop volumes of 2.4 to 8 ml (diameters of 16 to 25 mm) were formed in 5 to 15 s respectively (the
maximal pumping time was purposely restricted to 15 s according to Sect. \ref{drop_oscillations}). After each
microgravity phase, the water drop fell down under hypergravity and was passively absorbed by sponges covering
the vessel bottom.

\subsection{Drop Oscillations in g-Level Jitter}\label{drop_oscillations}

\begin{figure}[h]
  \includegraphics[width=\columnwidth]{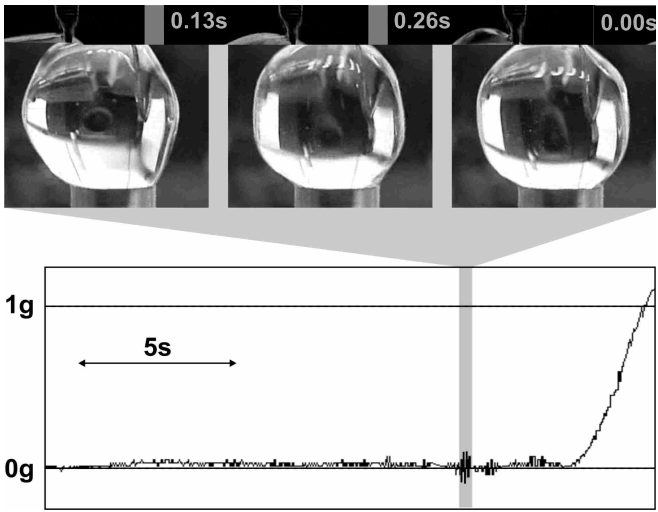}
  \caption{(\textit{top}) Water drop oscillations resulting from g-jitter superior to 0.05 g, as filmed by the standard
  video camera (Sect. \ref{general_setup}). (\textit{bottom}) Vertical component of the acceleration vector \textit{vs.} time}
  \label{fig_oscillation}
\end{figure}

During the microgravity phases provided by parabolic flights, the residual g-jitter can reach values of the order
of 0.02 g at typical frequencies of 1-10 Hz. In 10\% of the zero-g parabolas however, we recorded flight-based
deviations from the microgravity level larger than 0.02 g for durations longer than 0.2 s. They mostly occurred
approximately 3 s before the end of the zero-g phases. The injector tube being fixed to the aircraft reference,
those fluctuations induced observable multi-pole oscillations of the water drop, visible in Fig.
\ref{fig_oscillation}. Although the g-jitter spectrum contains eigenfrequencies of the drop \citep[see][for a
theoretical evaluation of the natural frequencies of captured drops]{Bauer 04}, no resonance was observable and
these oscillations always damped out rapidly within characteristic times of 0.5 to 1 s. We argue that the large
area of support attaching the drop to the tube (1.1 cm$^2$ for an outer diameter of 12 mm) was responsible for
this efficient attenuation, as the large porous foam ensures an inelastic rebound of the water drop. Due to the
strong g-jitter sometimes seen in the final 3 seconds of the microgravity phase, we limited the pumping time to
15 s, thus creating the most undisturbed spherical drops for the cavitation studies described in Sect.
\ref{cavitation part}.

We note that unwanted oscillations could also be partly avoided by increasing the viscosity of the fluid (e.g.
using a solution of glycerol/water), or by using a second injector tube stabilizing the top of the drop, as
realized by \citet{Treuner 95}. On the other hand, the observed multipole oscillations qualitatively compare to
the theoretical calculations of a harmonically excited captured drop by \citet{Bauer 04}, indicating another
potential application of this drop generation system.

\subsection{Drop Equilibirum on Injector Tube}\label{drop_equilibrium}

Depending on the drop volume and the injector tube diameter, the drop could be unstable and sometimes even
detached from the injector tube before the end of the microgravity phase. This leads to the following question:
Given a tube diameter, what is the maximum drop volume that is assured to remain attached on the tube under normal
unevittable g-jitter?

We answer this question by formulating a simplified stability criterion for the drop, based on its mechanical
equilibrium in the tube non-inertial frame. Only the force balance is considered, while the torques and
shear-induced drop deformations are neglected. In the frame of the tube, the inertial force acting on the drop
volume is opposed to its acceleration, $\bm{F_V} = -\rho V \bm{a}$, where $V$ is the drop volume, $\rho=1000$
kg m$^{-3}$ its mass density and $\bm{a}$ the residual acceleration (g-jitter) of the tube in microgravity. In
reaction to this volume force, the cohesive forces at the tube interface develop a counterforce $\bm{F_S} =
\bm{\sigma_S} S$, where $S$ is the tube surface and $\bm{\sigma_S}$ the surface force density along the
direction considered. The drop will be stable as long as the tube surface force compensates the volume force,
namely $\rho V a < \sigma_{S,max}S$, where $\sigma_{S,max}$ is the maximal value of the surface force density
that can be exerted by the tube interface (considered as an intrinsic property of the tube). Although
$\sigma_{S,max}$ is in principle direction-dependent, we treat it as an average over
all directions. During a standard parabola, the g-jitter reaches a maximum value $a_{max} = $ 0.02 g and the
stability criterion can be formulated as:

\begin{equation}\label{equ_equilibrium}
\frac{V}{S} < \frac{\sigma_{S,max}}{\rho a_{max}} \equiv c
\end{equation}
Since the critical parameter $c$ is a constant, the ratio $V/S$ dictates whether the drop remains stable during
the microgravity phase.

\begin{figure}[hb]
  \includegraphics[width=\columnwidth]{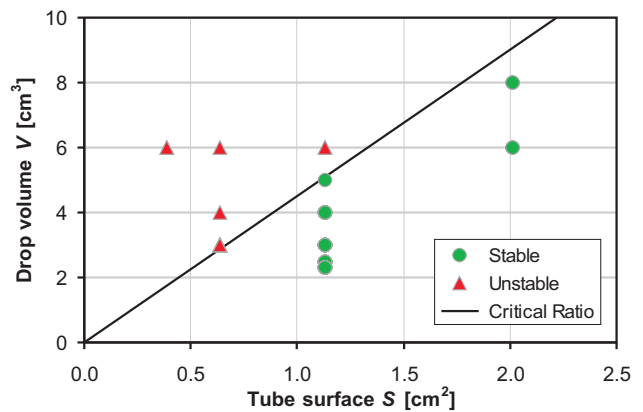}
  \caption{Diagram representing the different pairs $(V,S)$ of drop volume $V$ and tube contact surface $S$ used
  in the course of the experiment}
\label{fig_VvsS}
\end{figure}

This theoretical result could be verified in our microgravity experiment, which also allowed a phenomenologial
determination of $c$ as follows. For all 155 parabolas (summing over both flight campaigns), we collected the
drop volumes $V$ and tube surface areas $S$ and represented them as pairs $(S,V)$ in the diagram of Fig.
\ref{fig_VvsS}. In total, fourteen different $(S,V)$ configurations were probed, corresponding to the fourteen
data points. Each of these configurations was repeated in a variable number of parabolas. If during all of these
parabolas, the drops remained stable on the tube (i.e. never detached and hanged on the side of the tube), the
corresponding point was classified as stable (circles), otherwise as unstable (triangles).

In this representation, the critical parameter $c$ corresponds to the slope of a straight line passing through
the origin and separating the unstable from the stable regime; graphically, we find $c\sim4.5$ cm. This value has
to be taken as a safe stability limit, as no configuration with $V/S<4.5$ cm was unstable. For the 49
experimental cycles with $V/S>4.5$ cm, the average rate of drop detachment was about 25\%. Hence, configurations
with $V/S>4.5$ cm are not necessarily unstable, as rapid variations of the acceleration vector can prevent large
displacements of the drop or even bring it back to a stable state. Therefore the relation (\ref{equ_equilibrium})
should be exclusively taken as a \emph{stability} criterion, as its violation is only a necessary but not
sufficient condition for unstability.

We like to end this analysis with a note for additional physical insight: given the observed value $c\sim4.5$
cm, we can estimate the maximal surface force density $\sigma_{S,max}$ using (1). One finds
$\sigma_{S,max}\sim10$ N\,m$^{-2}$, which compares well with the cohesive force of water observed in droplets
suspended under their weights \citep[see][lecture L for further discussion on that matter]{Young}.

Stimulated by this reflection, we can find a physical interpretation for the length $c$ by noting that the ratio
$V/S$ corresponds to the height of a cylinder with base surface $S$ and volume $V$. Thus $c$ can be seen as the
maximally allowed height of a hypothetical water cylinder, such that can be sustained by the tube interface in a
weak gravitational field of 0.02 g.

\section{Generation and Study of Cavitation Bubble } \label{cavitation part}
\subsection{General Setup}\label{general_setup}

\begin{figure}[hb]
\centering
  \includegraphics[width=\columnwidth]{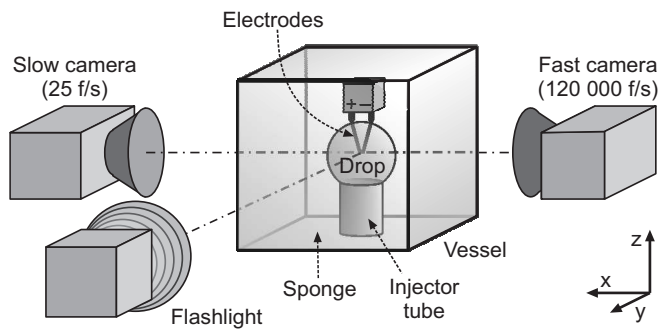}
  \caption{Configuration of the optical system surrounding the central components}
  \label{fig_central_components}
\end{figure}

Figure \ref{fig_central_components} shows the arrangement of the central components of the setup. The water drop,
the injector tube as well as the electrodes for the bubble generation were contained in a sealed transparent
cubic vessel (20 cm side length, 1 cm thickness) made of clear high-density polycarbonate (Lexan).

To record the fast bubble dynamics, we used a high-speed CCD camera (Photron Ultima APX, up to 120000 frames/s)
with a framing rate of 12500-24000 frames/s (depending on the temporal and spatial resolution desired), and an
exposure time of 5 $\mu$s. The record sequence had a duration of 11 ms, during which the required illumination
was provided by a perpendicularly placed high-power flashlight (Cordin Light Source Model 359). The slow water
drop formation was filmed with a standard video camera (Sony Camcorder DCR-TRV900, 25 frames/s).

\begin{figure}[ht]
  \includegraphics[width=\columnwidth]{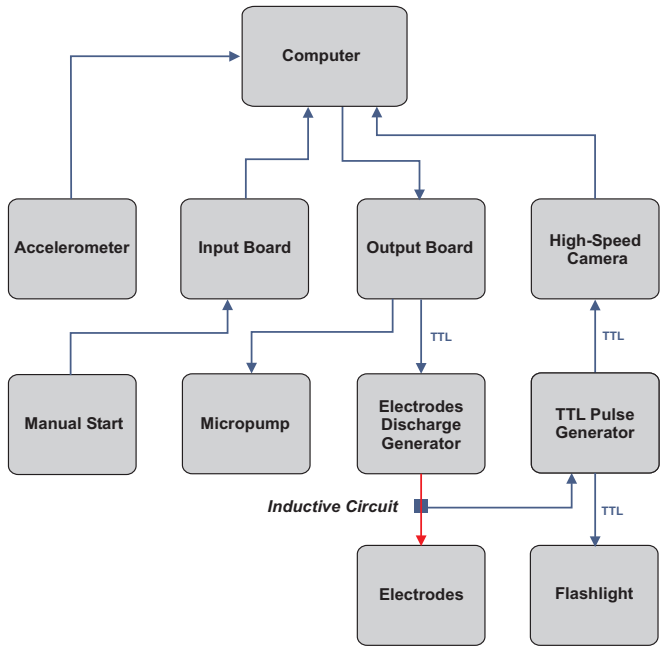}
  \caption{Flow diagram of the different functional units of the experiment in sequential order}
  \label{fig_flow_diagram}
\end{figure}

A diagram illustrating the links between the different functional units of the experiment as well as the order of
their activation in time is given in Fig. \ref{fig_flow_diagram}. A control computer program triggered the
micropump (drop creation), the electrodes discharge (bubble generation), the image recording and the flashlight
release. The interface between the computer signals (USB) and the manual start button, micropump and electrodes was
performed by standard input and output boards. The experimental sequence was automatically initiated at the
beginning of each flight parabola, as soon as a stable level of microgravity ($<$ 0.05 g) was reached for more
than 1 s. The gravity data was provided by a 100 Hz accelerometer (Memsic, 2125 Dual-axis Accelerometer) that
continuously recorded the gravity level during the whole flight. Immediately after the water drop formation, the
electrical discharge for bubble generation was released. Because of the low trigger time accuracy of the
discharge generator, we used the electrodes discharge pulse current to trigger the recording system through an
inductive response circuit. This precaution ensured that the recording system started at the instant of bubble
creation, in order to capture the entire bubble dynamics and subsequent phenomena. For the
exact synchronization of the high-speed camera and the flashlight, the induced signal from the electrodes was
split by a TTL pulse generator sending a pulse to both units simultanuously.

The experiment was designed to be redundant in case of failure of any unit: As an alternative for the accelerometer,
the experimental sequence could be started manually from an external button or directly from a computer key. If
none of these devices were functional, the electodes discharge generator could still be fired manually
instead of being triggered by the computer. Finally, the recording system could be triggered directly from the
electrodes inductive signal in case the TTL pulse generator were defective, but with a somewhat less accurate
synchronization. However, none of these off-nominal procedures had to be used over the 155 parabolas.

\begin{figure}[ht]
  \includegraphics[width=\columnwidth]{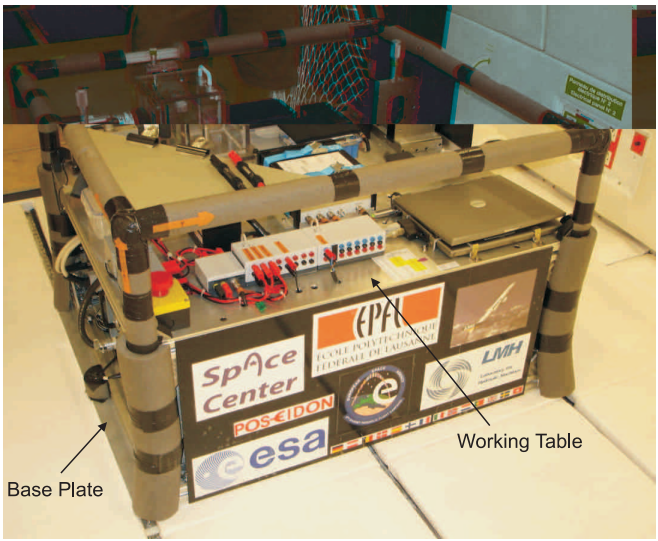}
  \caption{Picture of the two-level rack carrying the experiment}
  \label{fig_rack}
\end{figure}

The experimental setup was fixed on a single rack containing horizontal levels, as shown in Fig. \ref{fig_rack}.
The lower level (\emph{base plate}) was used to attach the experiment on the plane rails and to carry the heavy
units (high-power generators for electrodes and flash light, camera control unit), while the upper level
(\emph{working table}) carried the devices requiring in-flight manipulation.

\subsection{Spark-based Technique and Design of Electrodes}\label{electrodes}

The cavitation bubble was generated using the standard \emph{spark-based technique}, by which a spark released by
electrodes allows the conversion of electrostatic energy into a point-like plasma of high temperature and
pressure. The plasma immediately recombines and gives rise to a thermally growing bubble, which subsequently
collapses under the surrounding liquid pressure \citep{Chahine 95}. This standard \emph{spark-based technique}
was chosen instead of a laser-based technique because of the difficulty to focus a laser beam across a (variable)
spherical surface, and for aircraft security reasons.

\begin{figure}[h]
  \includegraphics[width=\columnwidth]{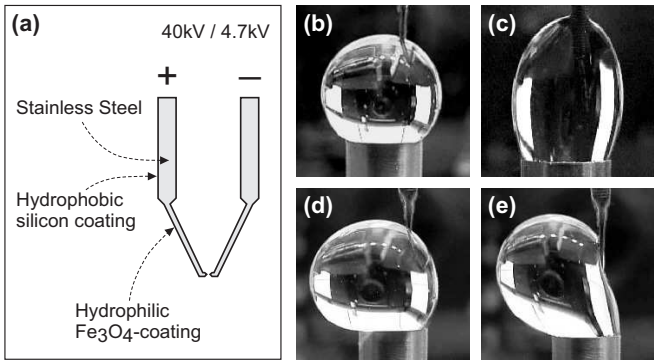}
  \caption{(a) Schematics of the electrodes with their coating (b) Quasi-spherical drop (b) Oblong
  attracted drop (c) Slightly repulsed drop (d) Strongly repulsed drop}
  \label{fig_electrodes_drops}
\end{figure}

To produce the discharge, a pair of electrodes was immersed in the water drop from above (Fig.
\ref{fig_electrodes_drops} (b-e)). The required electrostatic energy was stored into capacitors, whose fast
discharge was enabled by a triggered spark gap \citep[see][for more details]{Pereira 94}. These devices were
contained in the unit called \emph{Electrodes Discharge Generator} in Fig. \ref{fig_flow_diagram}. The
capacitor's voltage (4.7 kV) was high enough to induce the breakdown of water molecules and thereby create the
plasma between the electrodes. To vary the bubble size, the discharge capacitance could be stepwisely altered
between 30, 50, 100 and 200 nF. A minimum capacitance value of 30 nF was necessary to ensure water breakdown,
probably due to the potential drop occuring during the finite impedance fall-time of the spark gap.

The electrodes had a thicker part which was mounted on an axis controllable by micro-positioners, and thinner
ends to penetrate the drop (see Fig. \ref{fig_electrodes_drops} (a)). The latter had a thickness of the order of
0.5 mm with a spacing between the tips of about 1 mm. Using the positioners, the electrode tips could be
displaced in three dimensions, allowing their precise positioning inside the water drop.

In order to minimize the interaction of the electrodes with the water volume, their tips were coated with hydrophilic
ferrous oxyde Fe$_3$O$_4$ and their upper part covered with hydrophobic silicon, as illustrated in Fig.
\ref{fig_electrodes_drops} (a). Despite of the careful preparation of the electrodes, we observed significant
water drop distortions induced by electrode-water interaction. These distortions were strongly dependent on the
specific coating of the electrodes, particularly on the thickness of the silicon layer covering their thick upper
part. Without any silicon coating, the upper part of the electrodes was too hydrophilic, yielding an oblong
distortion of the water volume (Fig. \ref{fig_electrodes_drops} (c)). Conversely, additional silicon engendered water
repulsion by the electrodes (Fig. \ref{fig_electrodes_drops} (d,e)). Because of this effect, the electrodes position
had to be readjusted during the flight to control the direction of liquid jets emission following bubble collapse
(directed towards to and away from the closest free surface element, see Sect. \ref{jets and drop_stability}), so
that they did not escape from the high-speed camera field of view. Moreover, the drop deviations from a perfect
sphere hinder the measurement of quantitative parameters such as the curvature ratio between drop and cavity,
relative bubble eccentricity $\epsilon$ and relative bubble radius $\alpha$ (defined in Sect. \ref{jets and
drop_stability}). These parameters are nevertheless of importance for the analysis and modeling of the
observations. The above thus reveals the necessity to optimize the electrodes surface coating and to render the tips
finer in order to minimize such distortions.

\subsection{Discharge-Cavity Energy Relation}\label{discharge-cavity}

To characterize the efficiency of our spark-based system, we determined the fraction of electrical discharge
energy actually transformed into bubble energy. The electrical discharge energy $E_e$ is given by the stored
electrostatic energy of the capacitors. The hydrodynamic potential energy of the bubble $E_b$ relates to its
maximum radius $R_{b,max}$ (as achieved at the end of its growth) via $E_b = \frac{4\pi}{3}
R_{b,max}^3(p_\infty-p_v)$, where $p_\infty$ is the ambient static pressure and $p_v$ the water vapor pressure at
room temperature. In order to determine the relation $E_b(E_e)$ experimentally, we therefore measured the maximum
bubble radii $R_{b,max}$ for the different values of the capacitance.

To complement the flight measurements within the drops, we also carried out on-ground measurements inside water
flasks with additional values of the capacitance: 10 and 20 nF. For the measurements inside the drops, we
corrected the apparent $R_{b,max}$ using a model of optical refraction through a spherical surface (see appendix
\ref{optical distortion}). Care was taken to select bubbles sufficiently far from the drop surface to ensure the
validity of the refraction model as well as to avoid bubble shape disturbances. For each measured value of
$R_{b,max}$, the corresponding bubble energy $E_b(R_{b,max})$ was determined using the relation above, taking
into account the different ambient pressures on ground and inside the aircraft cabin, $10^5$ and $(8\pm0.2)10^4$
Pa respectively. Note that the cabin pressure contains an uncertainty owing to dynamic changes in the course of a
parabola (e.g. during throttle-back maneuvers), but we did not correct for these changes (between different
measurements) here and considered the pressure constant over the bubble lifetime. The mean values of $E_b$ are
plotted against $E_e$ in Fig. \ref{fig_energy_relation}. The errors on $E_b$ are the standard deviations of the
$E_b(R_{b,max})$ measurements, while the $E_e$ errors account for the standard precision of capacitance values of
about 5 $\%$. A weighted least-square linear regression (accounting for both errors) yielded the following
discharge-cavity energy relation: $E_b = 0.024 (\pm 0.005) E_e$.

\begin{figure}[h]\label{fig_energy_relation}
  \includegraphics[width=\columnwidth]{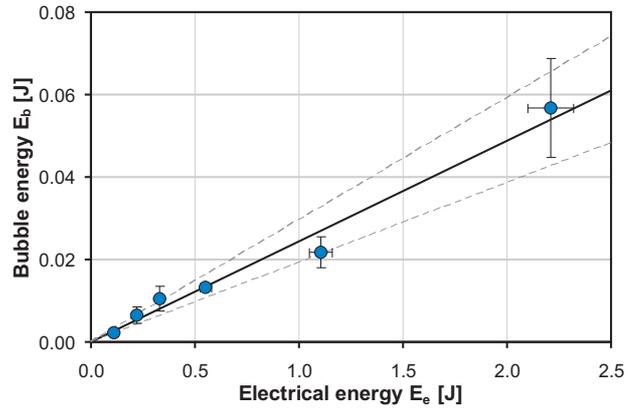}
  \caption{Mean bubble energy as a function of capacitors electrostatic energy.
  (\textit{solid line}) Weighted linear regression. (\textit{dashed line}) Regression uncertainties}
  \label{fig_energy_relation}
\end{figure}

The obtained relation reveals that about 2-3 \% of the capacitors electrostatic energy was transformed into
potential energy of the bubble, whereas the remaining energy was absorbed elsewhere. The latter fraction was
presumably dissipated in electrical losses and radiated in the form of a shock wave caused by the initial plasma
expansion \citep{Chahine 95}. To estimate the contribution of the shock wave proper, we used the method described
in \citet{F&S 06}. Explicitely, we assumed that most of the shock energy was converted into submillimetric
bubbles, arising from the excitation of impurities at the multiple passages of the shock confined within the
drop. By integrating the potential energies of all these bubbles (using the energy-maximum radius relation given
above), we found the fraction of shock wave energy to be roughly 4 \% \citep[cf][where it was found that the
shock wave energy is about 50-70 \% of the sum of the shock wave energy and bubble energy, confusingly referred
to as the "initial electrical energy"]{F&S 06}. By taking into account additional sources of energy loss such as
heat production and light emission \citep{Brenner 02, Leighton 94}, the actual energy deposited in the water by the electrodes
can be estimated to be of the order of 10 \%. The remaining 90 \% correspond to the initial energy fraction lost
in electrical Joule heating, electro-magnetic radiation, and plausible incomplete discharge of the capacitors (as
the electrodes voltage drops below the breakdown threshold of water). Although these values depend much on the
specific devices used, they are roughly consistent with previous estimations \citep{Chahine 95}.

\subsection{Liquid Jets and Drop Stability after Bubble Collapse}\label{jets and drop_stability}

Next to a free surface, a cavitation bubble undergoes a toroidal collapse emitting a high-speed \emph{microjet}
directed away from the local surface element, while a counterjet called \emph{splash} forms on the free-surface
\citep{Blake 81, Robinson 01, Pearson 04}. In the present drop experiment, two liquid jets were then seen to
escape the drop in opposite directions \citep{F&S 06}. We observed that the importance of these liquid jets
strongly depends on the bubble size and position within the drop. Furthermore, the collapsing bubble could induce
instabilities on the drop surface or even cause a burst of the entire drop, thus rendering the jets investigation
impossible. This motivated us to study the influence of the geometrical configuration of the bubble within the
drop on these particular phenomena.

\begin{figure}[ht]
\centering
  \includegraphics[width=\columnwidth]{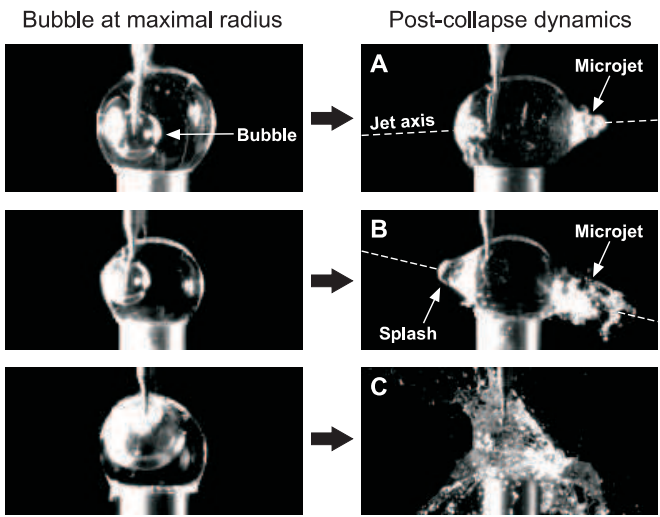}
  \caption{Three qualitative categories characterizing the strength of jets emission and the stability of the
  water drop and its surface after bubble collapse. (\textit{left column}) Instant of maximum bubble size, illustrating
  the relative bubble radius and its position within the drop. (\textit{right column}) Situation 4 ms after, illustrating
  the three post-collapse dynamics categories. Cat.A: The microjet escapes but the splash does not. Cat.B: The
  splash escapes and the microjet is more violent. Cat.C: The entire drop bursts}
\label{fig_categories}
\end{figure}

To characterize the jets emission and drop stability in function of the system's geometry, we
introduced three qualitative categories A, B, C, corresponding to the possible drop dynamics scenarios following
bubble collapse shown in Fig. ~\ref{fig_categories}.

The geometrical quantities dominating the post-collapse dynamics are (1) the drop size, expressed by the minimum
drop radius $R_{d,min}$ before bubble generation, (2) the bubble size, given by the maximum bubble radius
$R_{b,max}$, and (3) the bubble's radial position within the drop, characterized by the distance $d$ between drop
center and bubble center. But these three parameters are redundant, inasmuch as a uniform stretching of all
parameters reproduces a geometrically similar case. A natural choice of non-redundant dimensionless parameters
was then obtained by defining: the \emph{relative bubble radius} $\alpha\equiv R_{b,max}/R_{d,min}$ and the
\emph{bubble eccentricity} $\varepsilon\equiv d/R_{d,min}$.

For the 67 parabolas that qualified for this study (successful bubble generation, quasi-spherical drop, jets
within the high-speed camera field of view), the geometrical parameters $\alpha$ and $\epsilon$ were measured as
follows. Because of the drop deviation from spherical symmetry in presence of the electrodes (see section
\ref{electrodes}), the geometrical drop center could not be defined accurately in most parabolas. We therefore
adopted the natural convention to measure $R_{d,min}$ and $d$ along the axis of jets emission, being the relevant
physical axis for the dynamics of the system. As this axis was determined only after the jets emission, it was
exported to the first frames of the movies for the measurement of $R_{d,min}$ and $d$. To obtain the true
position (i.e. corrected for optical refraction through the drop's surface) of the bubble center on this axis,
the outer part of the electrode tips was prolongated inside the drop until its intersection with the jets axis.
Finally, $R_{b,max}$ was obtained from the linear regression between bubble energy and electrostatic energy
$E_b(E_e)$ (derived in Sect. \ref{discharge-cavity}), provided the corresponding value of the capacitance (and
the electrostatic energy thereof) and the bubble energy-maximum radius relation $E_b(R_{b,max})$.

\begin{figure*}[hbt]
\centering
  \includegraphics[width=17.6cm]{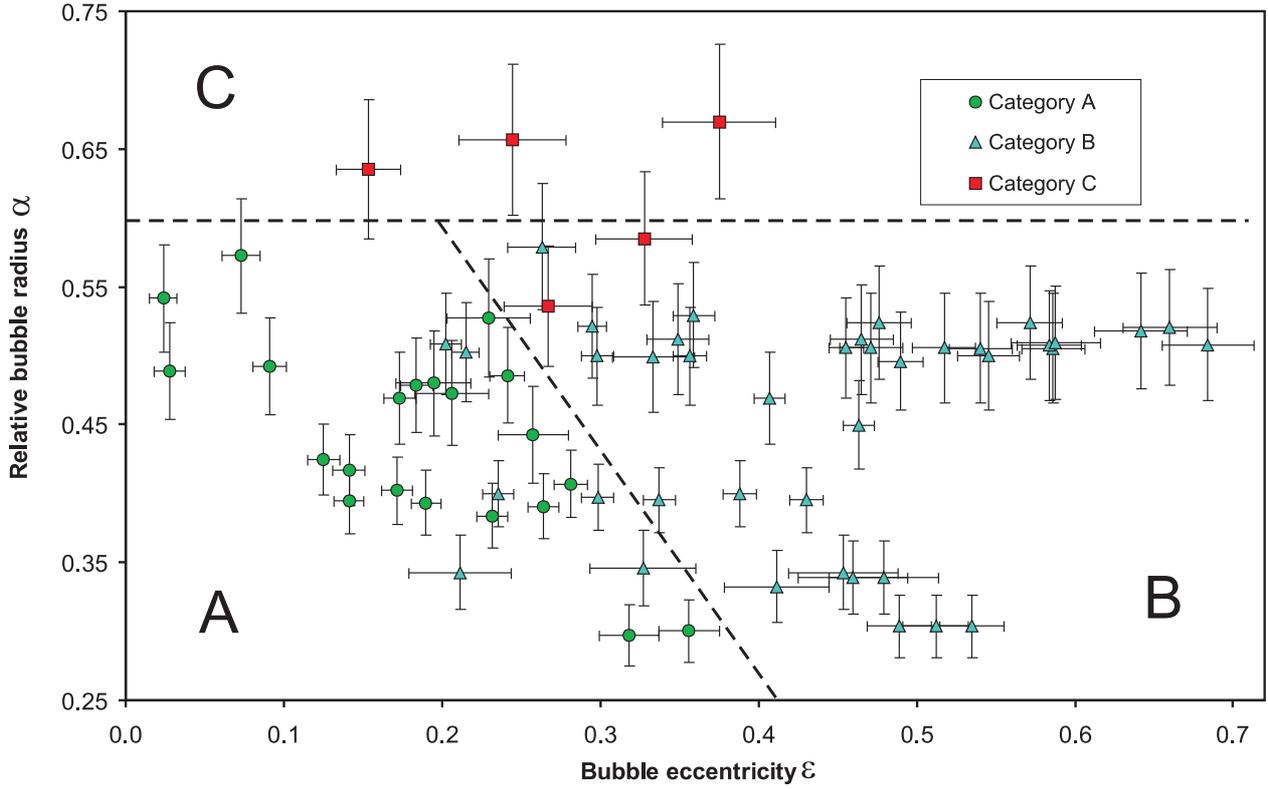}
  \caption{Representation of 67 parabolas in the parameter space
  \{bubble eccentricity $\epsilon$, relative bubble radius $\alpha$\}. The parabola points $(\epsilon, \alpha)$
  belonging to different categories A,B,C (introduced in Fig. \ref{fig_categories}) were plotted with different
  symbols, in order to identify associated parameter space domains. Straight lines roughly delimiting these
  domains were overdrawn (\textit{dashed}). The error bars for $\epsilon$ depend on the optical resolution (objective
  focal length), while the uncertainty on $\alpha$ was propagated from the uncertainty on the $E_b(E_e)$
  regression (section \ref{discharge-cavity})}
  \label{fig_alpha_vs_epsilon}
\end{figure*}

We then assigned each of these 67 parabolas to one of the three A-B-C categories, and represented them in the
dimensionless parameter space $\{\epsilon,\alpha\}$ shown in Fig. \ref{fig_alpha_vs_epsilon}. In that space, we
can roughly recognize three domains corresponding to the categories A, B and C, for which we have drawn straight
boundaries to guide the eye (dashed lines). It can be observed that for $0.3 < \alpha < 0.6$, increasing the
eccentricity leads to a transition from category A to category B. A similar transition occurs by increasing
$\alpha$ for $0.25 < \epsilon < 0.45$. Hence, the splash strength qualitatively increases both with $\alpha$ and
$\epsilon$, as larger values of these parameters may produce a stronger pressure peak between the bubble boundary
and the local free surface \citep[as depicted in the numerical simulations of][]{Robinson 01}. Notably, for
relative bubble radii above $\alpha \sim 0.6$, we assist to a burst of the entire water drop (cat. C). A closer
look at the five cases belonging to this category revealed that the drops remain essentially stable during bubble
growth and collapse, but explode immediately after, presumably under the effect of the strong shock wave radiated
by the rebounding bubble \citep{Shima 97}.

Consequently, as long as only the cavitation bubble life cycle is to be studied, the whole parameter space
$\{\epsilon,\alpha\}$ can be experimentally explored. However, to investigate the jets dynamics and surface
effects, the parameter space should be restricted to $\alpha<0.6$.

\section{Summary and Outlook}\label{summary}

We have implemented a simple injector tube system to generate centimetric, quasi-spherical captured drops in
microgravity. Mechanical considerations combined with observations revealed that the drop equilibrium is
essentially dictated by the ratio of the drop volume to the area of the tube interface. By studying the
equilibirum of all the probed configurations, we could estimate the maximum value of this volume-to-contact
surface ratio to ensure drop stability.

Single bubbles were generated by spark release between the tips of electrodes immersed
within the drop. A careful preparation of the electrode coating was shown to be necessary to avoid drop
distortion and deviation from spherical shape. By measuring the maximum bubble radii for various values of the
capacitance, we determined the fraction of the electrostatic discharge energy effectively converted into bubble
potential energy. In the remaining energy balance, we estimated the fraction of energy carried by the spark shock
wave, and lost in electrical Joule heating. Finally, we investigated the influence of the bubble position and
size on the strength of the liquid jets and the drop surface perturbations following bubble collapse. It was
found that the jets become more prominent as both the bubble eccentricity and size increase. In particular, for
relative bubble radii large enough, the whole drop occurs to burst immediately after bubble collapse.

Cavitation studies apart, our simple and low cost drop generator could offer other interesting applications. In
particular, we mention the study of capillary waves on a spherical free surface, as well as the modes and
frequencies of a harmonically-excited captured drop \citep[numerically simulated by][]{Bauer 04}. In the aim of
realizing two-phase flows in microgravity conditions, a similar drop generator could be used to study the process
of drop formation in flowing liquids, which could also be compared with theoretical models \citep{Kim 94}.

\begin{acknowledgements}
We gratefully acknowledge the \emph{European Space Agency (ESA)} for having offered the possibility to pursue
these experiments on parabolic flights. Furthermore, our gratitude is directed to the \emph{Swiss National
Science Foundation (SNSF)} and our private donator, the \emph{Swiss Space Industry Group (SSIG)}, who provided
the substantial basis of the whole research frame. We also thank several institutions of the \emph{\'{E}cole
Polytechnique F\'{e}d\'{e}rale de Lausanne (EPFL)}, the \emph{EPFL Space Center} in particular, for their support
in financial, practical and theoretical aspects. Finally, we adress a special thank to \emph{Claude Nicollier},
for having kindly offered his time for our support, and provided fruitful advices and experiences.
\end{acknowledgements}

\begin{appendix}
\section {Optical Distortion by a Liquid Drop}\label{optical distortion}

The image of the bubble inside the water drop is distorted due to optical refraction at the drop's surface. We
show that the apparent image is mostly too large, the linear stretching being approximately equal to the index of
refraction of water $\eta_{water}$.

\begin{figure}[h]
  \includegraphics[width=7.0cm]{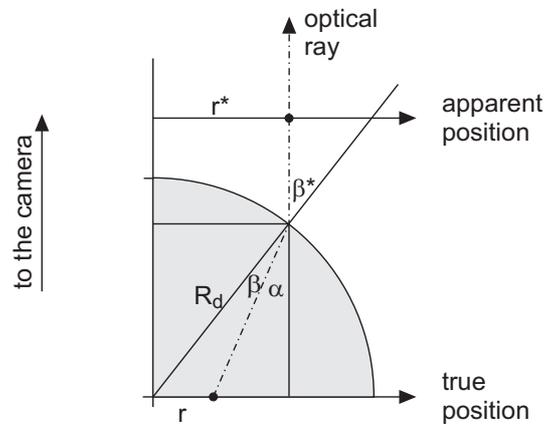}
  \caption{Model of optical refraction by a spherical drop. (\textit{gray zone}) Quarter of the drop's
cross-section. (\textit{dashed-dot line}) Optical ray}
  \label{fig_refraction}
\end{figure}

\begin{figure}[h]
  \includegraphics[width=7.5cm]{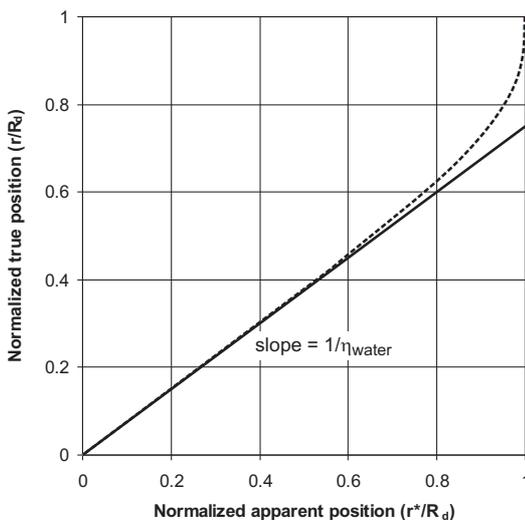}
  \caption{(\textit{solid line}) True position versus apparent position of a point inside a water sphere, observed perpendicularly to
  the straight line connecting it to the spheres center. (\textit{dashed line}) Linear expansion about $r=r^\ast=0$}
  \label{fig_true_vs_app}
\end{figure}

We consider a point located inside a spherical water drop at a distance $r$ from the drop center and placed
perpendicularly to the optical axis (Fig.\ \ref{fig_refraction}). From outside the drop, the point seems to lie
at an \emph{apparent position} $r^\ast$. The geometrical problem may be reduced by considering a two-dimensional
drop section parallel to the optical axis, containing both the drop center and the observed point. For
simplicity, only the relevant quarter of the drop section is sketched here. From Fig.\ \ref{fig_refraction} we
derive the geometrical relation
\begin{equation}
  r^\ast-r=\sqrt{R_d^2-{r^\ast}^2}\ tan(\alpha)=\sqrt{R_d^2-{r^\ast}^2}\ tan(\beta^\ast-\beta)
\end{equation}
where $sin(\beta^\ast)=r^\ast/R_d$. To eliminate $\beta$, Snell's law of refraction is applied,
\begin{equation}
  sin(\beta)=\frac{1}{\eta_{water}} sin(\beta^\ast)=\frac{r^\ast}{\eta_{water} R_d}
\end{equation}
After substitution and rearrangement, the following relation between $r$ and $r^\ast$ is obtained,
\begin{equation}
  r=r^\ast-\sqrt{R_d^2-{r^\ast}^2}\cdot tan\left(arcsin\frac{r^\ast}{R_d}-arcsin\frac{r^\ast}{\eta_{water} R_d}\right)
\end{equation}
Fig.\ \ref{fig_true_vs_app} (solid line) shows $r/R_d$ in function of $r^\ast/R_d$ for the particular index
$\eta_{water}=1.33$. Over the wide range $r^\ast/R_d<0.6$ this function is nearly linear obeying the simple law
(dashed line),
\begin{equation}
r^{\ast} = \eta_{water} r \qquad\text{(linear expansion about $r=r^\ast=0$)}
\end{equation}

\end{appendix}



\end{document}